\documentclass[fp,twocolumn]{jpsj3}
\usepackage{bm}
\usepackage{braket}
\usepackage{graphicx,color}
\usepackage{color}

\usepackage{ulem}
\usepackage{comment}

\title{Nonadiabatic Correction and Adiabatic Criteria \\ of Noninteracting Quantum Dot Systems}

\author{Masahiro Hasegawa$^1$ and Takeo Kato$^1$\thanks{kato@issp.u-tokyo.ac.jp}}
\inst{$^1$Institute for Solid State Physics, The University of Tokyo, Kashiwa, Chiba 277-8581, Japan} 

\abst{We theoretically study nonadiabatic corrections for charge pumping in a noninteracting electron model of a single-level quantum dot.
We derive a formula for the velocity limit of parameter driving to realize adiabatic pumping and illustrate its features in the wide-band limit.
We discuss the effect of the nonadiabatic corrections in terms of a typical velocity limit, which is defined by averaging the velocity limits on the parameter contour.
We also show that the typical velocity limit vanishes when the adiabaticity breakdown occurs in a quantum dot coupled to reservoirs with a band edge.
}

\begin{document}
\maketitle

\section{Introduction}
\label{sec:intro}

Adiabatic pumping, a transport phenomenon induced by quasi-static periodic driving of model parameters, has been investigated for a long time in various physical systems, since its concept was first proposed by Thouless~\cite{Thouless1983May}.
In the field of mesoscopic physics, electron transport induced by adiabatic pumping has been studied in a number of theoretical and experimental studies~\cite{Kouwenhoven1991Sep,Pothier1992Dec,Buttiker1993Jun,Buttiker94,Pretre1996Sep}.
Charge transferred per cycle of adiabatic pumping can be expressed in terms of the Berry curvature for non-interacting quantum systems~\cite{Brouwer1998Oct}, and a similar geometrical framework has been generalized to interacting electron systems~\cite{Aleiner1998Aug,Wohlman2002Apr,Brouwer05,Splettstoesser2005Dec,Splettstoesser2006Aug,Hernandez09,Reckermann10,Andergassen2010Jun,Calvo12,Pekola2013Oct,Hasegawa2017Jan,Aono04,Splettstoesser05,Sela06,Eissing16,Romero17,Hasegawa2018Mar}. 
Adiabatic pumping has also been examined in the context of quasi-static operations to characterize non-equilibrium states in nanoscale quantum heat engines~\cite{Benenti2017Jun,Yuge2013Nov,Nakajima2017Dec,Esposito2015Feb,Ochoa2018Feb,Chamon2011Apr,Juergens2013Jun,Whitney2014Apr,Shiraishi2016Oct}.

The geometric structure of adiabatic pumping is helpful for understanding non-equilibrium properties of nanoscale quantum engines.
However, in order to justify a geometric description of adiabatic pumping, its pumping protocol has to satisfy an adiabatic (or quasi-static) condition in the target systems.
In previous theoretical work, the adiabatic condition was not studied seriously and was roughly estimated intuitively or empirically.
However, in order to estimate the non-equilibrium properties of nanoscale engines, for example, the upper bound of efficiency of pumping, one has to evaluate the maximum velocity of parameter driving to satisfy the adiabatic condition because the parameter driving should be as fast as possible to increase the pumping power.
In particular, the upper limit of the driving velocity generally depends on the position in the parameter space, which is crucial to geometrical optimization of the pumping power~\cite{Hasegawa2020}.
This indicates that estimation of the adiabatic condition is not a simple problem.
Therefore, to be able to understand the non-equilibrium nature of nanoscale engines, the adiabatic condition should be defined in a systematic way.

\begin{figure}[tb]
    \centering
    \includegraphics[width=0.9\columnwidth]{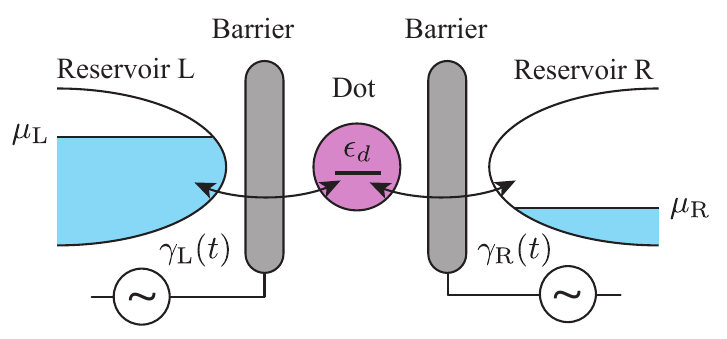}
    \caption{(Color online) Schematics of our model.
    A quantum dot is coupled to two electron reservoirs, L and R, whose chemical potentials are denoted by $\mu_{\rm L}$ and $\mu_{\rm R}$, respectively.
    The dot-reservoir couplings denoted by $\gamma_{\rm L}(t)$ and $\gamma_{\rm R}(t)$ are varied periodically by applying ac modulation to the potential barriers.
    }
    \label{fig:model}
\end{figure}

To this end, we theoretically derive a detailed condition for adiabaticity by taking the dependence of the parameter driving contour into account in this work. 
As a minimum setup for adiabatic pumping, we choose a non-interacting electron system composed of a quantum dot and two reservoirs as shown in Fig.~\ref{fig:model} and consider adiabatic pumping induced by periodic modulation of the dot-reservoir coupling.
Our formula constitutes a general procedure to calculate the upper limit of the driving velocity as a function of parameters and can be used to detect bottleneck regions in the parameter space at which the limit velocity is suppressed.
We examine the adiabatic condition by analyzing the leading nonadiabatic corrections for a quantum dot system coupled to reservoirs with a constant density of states by assuming an infinite bandwidth.
Then we examine a quantum dot coupled to reservoirs with band edges and discuss the problem of adiabaticity breakdown~\cite{Hasegawa2019b} in terms of our formulation of the adiabatic condition.

This paper is organized as follows.
In Sect.~\ref{sec:model}, we introduce a model of a quantum dot and formulate the transferred charge per cycle of parameter driving in terms of Keldysh Green's functions.
In Sect.~\ref{sec:AdiabaticApproximation}, we derive a nonadiabatic correction for the transferred charge and derive an adiabatic condition for the velocity of parameter driving.
In Sect.~\ref{sec:VelocityLimitInWidebandLimit}, we evaluate the adiabatic condition for a quantum dot coupled to reservoirs with a constant density of states in the wide-band limit.
In Sect.~\ref{sec:AdiabaticityBreakDownInBandEdgeSystem}, we examine adiabaticity breakdown in a quantum dot coupled to the reservoirs with band edges.
We summarize our results in Sect.~\ref{sec:Summary}.
The appendices present details on certain relevant equations.

Throughout this paper, we employ the unit of $\hbar = 1$ and denote the electron charge as $e$ ($<0$).

\section{Model}
\label{sec:model}

\subsection{Model Hamiltonian}

We consider a noninteracting electron model of a single-level quantum dot coupled to two electric reservoirs (see Fig.~\ref{fig:model}) described by the Hamiltonian:
\begin{align}
    H(t) = \epsilon_d d^{\dagger} d + \sum_{r,k} \left[ \epsilon_k c_{rk}^{\dagger} c_{rk} + \gamma_r (t) (d^{\dagger} c_{rk} + c_{rk}^{\dagger} d) \right] . \label{eqn:Hamiltonian}
\end{align}
Here, $d$ and $d^{\dagger}$ are the annihilation and creation operators of an electron in the quantum dot with an energy level $\epsilon_d$, while $c_{rk}$ and $c_{rk}^{\dagger}$ are those of electrons with a wavenumber $k$ and an energy $\epsilon_k$ in the reservoir $r = \mathrm{L}, \mathrm{R}$, respectively.
The time-dependent coupling constant between the dot and the reservoir $r$ is denoted as $\gamma_r (t)$.
We assume that the reservoir temperature is zero and set the reservoir electrochemical potentials as $\mu_r$.
Then, the distribution function of the reservoir $r$ can be described as
\begin{align}
    f_r(\omega) = \Theta (\mu_r - \omega),
\end{align}
where $\Theta(\omega)$ is the Heaviside step function.

Throughout this paper, we assume that both reservoirs have the same continuous density of states $\rho(\omega)$.
In the wide-band limit, the strength of the dot-reservoir coupling is described by a time-dependent linewidth, as
\begin{align}
    \Gamma_r(t) = 2\pi \rho_{0} |\gamma_r(t)|^2 ,
\end{align}
where $\rho$ is the density of states at the Fermi energy.
In Sect.~\ref{sec:AdiabaticityBreakDownInBandEdgeSystem}, where we discuss the breakdown of adiabaticity, we need to consider the dependence of the density of states carefully and use a time-dependent linewidth, defined as 
\begin{align}
    \Gamma_r(\omega;t) = 2 \pi \rho(\omega) |\gamma_r(t)|^2 .
\end{align}

\subsection{Transferred charge in one cycle}

We consider charge pumping induced by a time-dependent dot-reservoir coupling.
The charge transferred from the reservoir $r$ into the quantum dot in one pumping cycle, denoted by $\delta Q_r$, is calculated as
\begin{align}
    \delta Q_r &= \int_{0}^{T} dt \, \Braket{I_r(t)}, \\
    I_r(t) &= -e \frac{d}{dt} \left[ \sum_k c_{rk}^{\dagger}(t) c_{rk}(t) \right] \nonumber \\
    &= (-i) e  \gamma_r(t) \left( d^{\dagger}(t)c_{rk}(t) - c_{rk}^{\dagger}(t)d(t) \right),
\end{align}
where $T$ is the pumping period.
The average of the charge current operator can be expressed in terms of Keldysh Green's functions as
\begin{align}
    \Braket{I_r(t)} &= 2e \int dt_1 \ \mathrm{Re} \biggl[ G^R(t,t_1) \Sigma_r^<(t_1,t) \nonumber \\
    & \biggl. \hspace{10mm} +  G^<(t,t_1) \Sigma_r^A(t_1,t) \biggr] .
\end{align}
Here, $G$ and $\Sigma$ denote the Green's function and the one-particle-irreducible self-energy, respectively, and the superscripts, $R$, $A$, and $<$ indicate the retarded, advanced, and lesser components (see Appendix \ref{sec:KeldyshGreensFunction}).

\section{Adiabatic pumping and nonadiabatic correction}
\label{sec:AdiabaticApproximation}

\subsection{Average-time approximation}

To calculate the charge pumped in one cycle in the slow driving region, we employ the adiabatic approximation, that is called the average-time approximation in Ref.~\citen{Splettstoesser2005Dec}.
In this approximation, the time-dependent Hamiltonian is expanded up to the first order at frozen time $t$:
\begin{align}
    H(t^{\prime}) &\simeq H(t) + (t^{\prime}-t) \frac{d H(t)}{d t} \nonumber \\
    &=  H(t) + (t^{\prime}-t) \sum_r F_r \frac{d \gamma_r(t)}{dt},
\end{align}
where $F_r = \partial H/\partial \gamma_r$ is the response operator.
Pumped charge in one cycle is approximated into two terms,
\begin{align}
    \delta Q_r &\simeq \delta Q_r^{(0)} + \delta Q_r^{(1)} , \\
    \delta Q_r^{(a)} &= \int_{0}^{T} dt \ \Braket{I_r^{(a)}(t)}_{t}, \quad (a=0,1),  \label{eqn:0thAnd1stOrderCorrection}
\end{align}
where $\Braket{\cdot}_{t}$ denotes the steady-state ensemble average described by the Hamiltonian at frozen time $t$.
The current operators, $I_r^{(0)}$ and $I_r^{(1)}$, are calculated as
\begin{align}
    I_r^{(0)}(t) &= I_r(t) , \label{eqn:0thOrderOperator} \\
    I_r^{(1)}(t) &= \sum_{r_1} \int dt_1 \ \mathcal{T} \left[ I_r(t) F_{r_1} (t_1) \right] (t_1-t) \frac{d \gamma_{r_1}(t)}{dt}. \label{eqn:1stOrderOperator}
\end{align}
Here, $\mathcal{T}$ is the time-ordering operator.
The former term, $\delta Q_r^{(0)}$, is referred to as the steady-state term, because it is just the integral of the steady-state charge current in one cycle.
The latter term, $\delta Q_r^{(1)}$, is referred to as the adiabatic pumping term, because it is independent of the pumping period $T$, while the steady state contribution is of order $O(T)$.
This term can be interpreted in a geometrical manner by substituting Eq. (\ref{eqn:1stOrderOperator}) into Eq. (\ref{eqn:0thAnd1stOrderCorrection}) as
\begin{align}
    \delta Q^{(1)}_{r} &= \int_{0}^{T} dt \ \sum_{r_1} A_{r,r_1}(t) \frac{d \gamma_{r_1}(t)}{dt} , \label{eqn:1stOrderGeometric} \\
     A_{r,r_1}(t) &= \int dt_1 \, \langle {\cal T}[I_r(t)F_{r_1}(t_1)] \rangle (t_1-t).
\end{align}

\subsection{Adiabatic expansion}

In obtaining the pumped charge for adiabatic pumping in the previous subsection, the higher-order corrections, such as, $\frac{d^2 H(t)}{d t^2}$ and $\left( \frac{d H(t)}{d t} \right)^2$, were neglected.
In other words, adiabatic charge pumping is realized under the condition that these higher-order corrections are small enough compared to the adiabatic pumping term.
Here, in order to see the details of the condition for adiabatic pumping, we derive the higher-order corrections explicitly.

The pumped charge can be generally written as a series expansion with respect to the inverse of the pumping period, $1/T$:
\begin{align}
    \delta Q_r &= \sum_{n=0}^{\infty} \delta Q_r^{(n)} \\
    \delta Q_r^{(n)} &= \int_{0}^{T} dt \ \Braket{I_r^{(n)}(t)}_{t} \label{eqn:nthOrderCorrection}
\end{align}
Here, $\delta Q^{(n)}$ is of order $O((1/T)^{n-1})$ and $I^{(n)}_r(t)$ is referred to as the $n$th-order current operator.
The zeroth- and first-order current operators correspond to Eqs.~(\ref{eqn:0thOrderOperator}) and (\ref{eqn:1stOrderOperator}), respectively.

The second-order current operator is calculated as
\begin{align}
    I^{(2)}_r(t) = I^{(2,1)}_r(t) + I^{(2,2)}_r(t) \label{eqn:2ndOrderOperatorDef},
\end{align}
\begin{align}
    I^{(2,1)}_r(t) &= \frac{1}{2} \sum_{r_1} \int dt_1 \ \mathcal{T} \left[ I_r(t) F_{r_1}(t_1) \right] \nonumber \\
    & \hspace{10mm} \times (t_1-t)^2 \frac{d^2 \gamma_{r_1}(t)}{dt^2} ,
\end{align}
\begin{align}
    I^{(2,2)}_r(t) &= \frac{1}{2} \sum_{r_1,r_2} \int dt_1 dt_2 \ \mathcal{T} \left[ I_r(t) F_{r_1}(t_1) F_{r_2}(t_2) \right] \nonumber \\
    &\hspace{10mm} \times (t_1-t)(t_2-t) \frac{d \gamma_{r_1}(t)}{dt} \frac{d \gamma_{r_2}(t)}{dt} . \label{eqn:2ndOrderOperatorDetail2}
\end{align}
Substituting Eqs.~(\ref{eqn:2ndOrderOperatorDef})-(\ref{eqn:2ndOrderOperatorDetail2}) into Eq.~(\ref{eqn:nthOrderCorrection}), the second-order correction can be expressed in quadratic form,
\begin{align}
    \delta Q_r^{(2)} &= \int_{0}^{T} dt \ \Biggl( \sum_{r_1} B^{(1)}_{r,r_1}(t) \frac{d^2 \gamma_{r_1}(t)}{dt^2} \nonumber \\
    &\hspace{10mm} + \sum_{r_1,r_2} B^{(2)}_{r,r_1r_2}(t) \frac{d \gamma_{r_1}(t)}{dt}\frac{d \gamma_{r_2}(t)}{dt} \Biggr) \\
    B_{r,r_1}^{(1)}(t) &= \int dt_1 \, \langle {\cal T}[I_r(t)F_{r_1}(t_1)]\rangle (t_1-t)^2, \\
    B_{r,r_1r_2}^{(2)}(t) &= \int dt_1 dt_2 \, \langle {\cal T}[I_r(t)F_{r_1}(t_1)F_{r_2}(t_2)] \rangle \nonumber \\
    & \hspace{10mm} \times (t_1-t)(t_2-t). 
\end{align}
Using integration by parts, the second-order correction becomes
\begin{align}
    & \braket{I_r^{(2)}(t)} = \sum_{r_1,r_2} B_{r,r_1r_2}(t) \frac{d \gamma_{r_1}(t)}{dt}\frac{d \gamma_{r_2}(t)}{dt} . \label{eqn:2ndOrderGeometric}, \\
    & B_{r,r_1r_2}(t) = B^{(2)}_{r,r_1r_2}(t) - \frac{\partial B^{(1)}_{r,r_1}(t)}{\partial \gamma_{r_2}}.
\end{align}

The third-order current operator is calculated in the same way:
\begin{align}
    I^{(3)}_r(t) &= I^{(3,1)}_r(t) + I^{(3,2)}_r(t) + I^{(3,3)}_r(t) \label{eqn:3rdOrderOperatorDef},\\
    I^{(3,1)}_r(t) &= \frac{1}{3!} \sum_{r_1} \int \! dt_1 \, \mathcal{T} \left[ I_r(t) F_{r_1}(t_1) \right] \nonumber \\
    & \times (t_1-t)^3 \frac{d^3 \gamma_{r_1}(t)}{dt^3} , \\
    I^{(3,2)}_r(t) &= \frac{1}{2!}  \sum_{r_1,r_2} \int \! dt_1 \, \mathcal{T} \left[ I_r(t) F_{r_1}(t_1) F_{r_2}(t_2) \right] \nonumber \\
    & \times (t_1-t)^2 (t_2-t) \frac{d^2 \gamma_{r_1}(t)}{dt^2} \frac{d \gamma_{r_2}(t)}{dt} , \\
    I^{(3,3)}_r(t) &= \frac{1}{3!} \sum_{r_1,r_2,r_3} \int \! dt_1 \, \mathcal{T} \left[ I_r(t) F_{r_1}(t_1) F_{r_2}(t_2) F_{r_3}(t_3) \right] \nonumber \\
    & \hspace{-9mm} \times (t_1-t) (t_2-t) (t_3-t) \frac{d \gamma_{r_1}(t)}{dt} \frac{d \gamma_{r_2}(t)}{dt} \frac{d \gamma_{r_3}(t)}{dt} .
\end{align}
Using integration by parts, the third-order correction can be summarized into two terms:
\begin{align}
    & \braket{I_r^{(3)}(t)} = \braket{I_{r}^{(3v)}(t)} + \braket{I_{r,a}^{(3a)}(t)} ,\\
    & \braket{I_{r}^{(3v)}(t)} =\! \! \sum_{r_1,r_2,r_3} \! \! C_{r,r_1r_2r_3}(t) \frac{d \gamma_{r_1}(t)}{dt} \frac{d \gamma_{r_2}(t)}{dt} \frac{d \gamma_{r_3}(t)}{dt}, 
    \label{eqn:3rdOrderGeometricv} \\
    & \braket{I_{r}^{(3a)}(t)} = \sum_{r_1,r_2} D_{r,r_1r_2}(t) \frac{d^2 \gamma_{r_1}(t)}{dt^2} \frac{d \gamma_{r_2}(t)}{dt} . \label{eqn:3rdOrderGeometric}
\end{align}
The first term, $\langle I_{r}^{(3v)}(t)\rangle$, consists only of the first derivative of the driving parameter, whereas the second term, $\langle I_{r}^{(3a)}(t) \rangle$, includes the second derivative of the driving parameter.
In general, we can derive higher-order corrections by using a similar procedure for the $n$th-order operator, defined as
\begin{align}
    I^{(n)}_r(t) = \int dt_1 \cdots dt_n \ \mathcal{T} \left[ I_r(t)  F_n(t_{1},\cdots,t_{n})  \right] ,
\end{align}
\begin{align}
    F_n(t_{1},\cdots,t_{n}) = \prod_{m=1}^{n} \sum_{l_m} \sum_{r^{\prime}} \frac{\partial H}{\partial \gamma_{r^{\prime}}}(t_m) \left(\frac{d }{dt}\right)^{l_m} \gamma_{r^{\prime}}(t) ,
\end{align}
though their expressions become much more complex.

\subsection{Adiabatic condition}
\label{sec:AdiabaticCondition}

We define the driving velocity of the parameters as
\begin{align}
{\bm v}(t) = (v_{\rm L}(t),v_{\rm R}(t)) = \biggl( \frac{d\gamma_{\rm L}}{dt},\frac{d\gamma_{\rm R}}{dt}\biggr) .
\end{align}
To write down the adiabatic condition, it is convenient to introduce the normalized velocity, defined as
\begin{align}
\tilde{\bm v}(t) =  (\tilde{v}_{\rm L}(t),\tilde{v}_{\rm R}(t)) = \frac{{\bm v}(t)}{|{\bm v}(t)|}.
\end{align}

Let us discuss how the driving velocity is restricted in order to realize adiabatic pumping.
Charge pumping can be regarded as `adiabatic' when the higher-order (nonadiabatic) corrections of the charge current are much smaller than the adiabatic current:
\begin{align}
    | \langle {I_r^{(2)}(t)} \rangle|, | \langle {I_r^{(3v)}(t)} \rangle|, | \langle {I_r^{(3a)}(t)} \rangle|, \cdots \ll | \langle I_r^{(1)}(t) \rangle| .
\end{align}
Note that this condition is always satisfied in the slow-driving limit of ${\cal T}\rightarrow \infty$ since $\langle I_r^{(n)}(t)\rangle$ is proportional to $(1/T)^{n-1}$.
For a finite speed of driving, we introduce a small tolerance factor $\eta$ ($\ll 1$) and describe the extent to which the ratio of the higher-order current corrections to the adiabatic current is allowed as follows:
\begin{align}
|\langle I_r^{(2)}(t)\rangle | \le \eta |\langle I_r^{(1)}(t)\rangle| , 
\label{eq:condI2}\\
|\langle I_r^{(3v)}(t)\rangle | \le \eta |\langle I_r^{(1)}(t)\rangle| ,
\label{eq:condI3v}
\\
|\langle I_r^{(3a)}(t)\rangle | \le \eta |\langle I_r^{(1)}(t)\rangle| .
\label{eq:condI3a}
\end{align}
We call these the adiabatic conditions.

The adiabatic condition for the second-order correction, Eq.~(\ref{eq:condI2}), can be rewritten using Eqs.~(\ref{eqn:1stOrderGeometric}) and (\ref{eqn:2ndOrderGeometric}) as
\begin{align}
|v_r(t)| \le v^{(2)}_{{\rm lim},r}(t), 
\end{align}
where the `velocity limit' $v^{(2)}_{{\rm lim},r}(t)$ is given as
\begin{align}
& v_{{\rm lim},r}^{(2)}(t) = \eta \frac{|A_r(t)|}{|B_r(t)|}, \label{eqn:v2Def}
 \\
& A_r(t) = \sum_{r_1} A_{r,r_1}(t) \tilde{v}_{r_1}(t), \\
& B_r(t) = \sum_{r_1,r_2} B_{r,r_1,r_2}(t) \tilde{v}_{r_1}(t) \tilde{v}_{r_2}(t) .
\end{align}
In a similar way, the adiabatic condition for the third-order correction, Eq.~(\ref{eq:condI3v}), can be rewritten as 
\begin{align}
    & |v_{r}(t)| \le v_{\mathrm{lim},r}^{(3)}(t), \\
    & v_{\mathrm{lim},r}^{(3)}(t) = \sqrt{\eta\frac{| A_r(t)|}{|C_r(t)|} } , \label{eqn:v3Def} \\
    & C_r(t) = \sum_{r_1,r_2,r_3} C_{r,r_1r_2r_3}(t) \tilde{v}_{r_1}(t) \tilde{v}_{r_2}(t) \tilde{v}_{r_3}(t).
\end{align}
The last condition, Eq.~(\ref{eq:condI3a}), imposes a restriction on the acceleration of the pumping parameter.
For simplicity, we will not discuss this condition and instead assume that the driving velocity is nearly constant (see Appendix~\ref{sec:ConditionAcceleration}).

Combining these conditions, we obtain
\begin{align}
    v(t) \leq \mathrm{min} \left[ v_{\mathrm{lim},r}^{(2)}(t), v_{\mathrm{lim},r}^{(3)}(t), \cdots \right] .
\end{align}
One might guess that, as long as one considers slow driving, the strictest condition is imposed by a second-order correction, since the third- and higher-order corrections are much smaller than the second-order one in the adiabatic expansion.
Although this is true in most cases, there is an exception:
When the system is in equilibrium, that is, when all the reservoirs have the same temperature and chemical potential, the second-order correction is always zero for any driving of $\gamma_r(t)$
(this exception will be discussed in the next subsection).
In this case, the velocity limit $v^{(2)}_{{\rm lim},r}(t)$ becomes infinite, and there is no restriction on the pumping velocity.
Therefore, the third correction should be included in the estimate of the velocity limit.

\subsection{Disappearance of the second-order correction}
\label{sec:The2ndOrderCorrectionDisappearance}

Comparing the charge transfer induced by a driving operation and its time-reversed operation, one can see that the $2n$-th-order correction disappears when the system is in equilibrium.
Let us consider a driving protocol $C$ defined in the parameter space, such as,
\begin{align}
    C=\{ (\gamma_L(t),\gamma_R(t))| t \in [0,T] \} ,
\end{align}
and its time-reversed protocol $\bar{C}$, such as,
\begin{align}
    \bar{C}=\{ (\gamma_L(T-t),\gamma_R(T-t))| t \in [0,T] \} .
\end{align}
In general, the amount of charge transfer induced by $C$ and $\bar{C}$ can vary, when steady charge current flows between the reservoirs or an external magnetic field is applied.
This indicates that when the system is in equilibrium, that is, when the temperature and chemical potential are the same among all the reservoirs, there is no steady charge current and the amount of charge transfer induced by $C$ and $\bar{C}$ should satisfy the following relation:
\begin{align}
    \delta Q_r[C] = - \delta Q_r[\bar{C}]. \label{eqn:AntiSymmetricInEq}
\end{align}
Here, $\delta Q [C]$ denotes the amount of charge transfer under the driving protocol $C$.
Let us consider the adiabatic expansion of $\delta Q[C]$ and $\delta Q[\bar{C}]$.
\begin{align}
    \delta Q_r [C] = \sum_{n=0} \delta Q^{(n)}_r[C], \\
    \delta Q_r [\bar{C}] = \sum_{n=0} \delta Q^{(n)}_r[\bar{C}]
\end{align}
Focusing on the first-order corrections (adiabatic pumping terms), they satisfy an anti-symmetric relation,
\begin{align}
    \delta Q^{(1)}_r [\bar{C}] &= \int_{0}^{T} dt \ \sum_{r_1} A_{r,r_1}(T-t) \frac{d \gamma_{r_1}(T-t)}{dt} \nonumber \\
    &= -\int_{0}^{T} dt \ \sum_{r_1} A_{r,r_1}(t) \frac{d \gamma_{r_1}(t)}{dt} \nonumber \\
    &= - \delta Q^{(1)}[C] ,
\end{align}
while the second-order corrections satisfy a symmetric relation,
\begin{align}
    &\delta Q^{(2)}_r [\bar{C}] \nonumber \\
    &= \int_{0}^{T} dt \sum_{r_1,r_2} B_{r,r_1r_2}(T-t) \frac{d \gamma_{r_1}(T-t)}{dt} \frac{d \gamma_{r_2}(T-t)}{dt} \nonumber \\
    &= \int_{0}^{T} dt \sum_{r_1,r_2} B_{r,r_1r_2}(t) \frac{d \gamma_{r_1}(t)}{dt} \frac{d \gamma_{r_2}(t)}{dt} = \delta Q^{(2)}_r[C] .
\end{align}
In general, the $(2n)$-th-order corrections and $(2n+1)$-th-order corrections satisfy symmetric and anti-symmetric relations, respectively,
\begin{align}
    \delta Q^{(2n)}[C] &= \delta Q^{(2n)}[\bar{C}] , \\
    \delta Q^{(2n+1)}[C] &= -\delta Q^{(2n+1)}[\bar{C}] . \label{eqn:EvenOrderSymmetric}
\end{align}
Combining Eq.~(\ref{eqn:AntiSymmetricInEq}) and Eq.~(\ref{eqn:EvenOrderSymmetric}), we can conclude that the $(2n)$-th-order correction should be zero for any driving protocols when the system is in equilibrium.
In this case, the nonadiabatic correction starts with the third-order correction $\delta Q^{(3)}$, so the velocity limit is imposed by the third-order correction, not the second-order correction.

\subsection{Typical velocity limit}

As shown in Eqs. (\ref{eqn:v2Def}) and (\ref{eqn:v3Def}), the velocity limit may drop to zero under certain conditions, which is known as adiabaticity breakdown.
There are two cases in which the velocity limit is zero:
(i) the adiabatic pumping term (numerator of Eqs.~(\ref{eqn:v2Def}) and (\ref{eqn:v3Def})) is zero;
(ii) the nonadiabatic corrections (denominator of Eqs. (\ref{eqn:v2Def}) and (\ref{eqn:v3Def})) are infinitely large.
The former case is not serious for adiabaticity, because the nonadiabatic corrections are finite and one can pass through such adiabaticity breaking points in finite velocity as long as the nonadiabatic corrections are negligible compared to the adiabatic pumping term in the whole cycle.
We refer to this case as `non-serious adiabaticity breakdown'.
On the other hand, the latter case is serious, because the nonadiabatic corrections diverge, so one should not pass through such points at any finite velocity.
We refer to this case as `serious adiabaticity breakdown'.

Unfortunately, it is difficult to distinguish serious and non-serious adiabaticity breakdown only by the velocity limit itself, and it is necessary to define an indicator that can distinguish them.
Here, we define a `typical' velocity limit as such an indicator,
\begin{align}
    \bar{v}_{\mathrm{lim},r}^{(2/3)} = \int_{0}^{T} dt \ v_{\mathrm{lim},r}^{(2/3)}(t) p^{(2/3)}(t) ,
\end{align}
where $p^{(2)}(t)$ and $p^{(3)}(t)$ are weight functions describing how large the nonadiabatic corrections are on the driving contour:
\begin{align}
    p^{(2)}(t) = \frac{\left| \sum_{r_1,r_2} B_{r,r_1r_2}(t) \tilde{v}_{r_1}(t) \tilde{v}_{r_2}(t) \right|}{\int_{0}^{T} dt \left| \sum_{r_1,r_2} B_{r,r_1r_2}(t) \tilde{v}_{r_1}(t) \tilde{v}_{r_2}(t) \right|},
\end{align}
\begin{align}
    p^{(3)}(t) = \frac{\left| \sum_{r_1,r_2,r_3} C_{r,r_1r_2r_3}(t) \tilde{v}_{r_1}(t) \tilde{v}_{r_2}(t) \tilde{v}_{r_3}(t) \right|}{\int_{0}^{T} dt \left| \sum_{r_1,r_2,r_3} C_{r,r_1 r_2 r_3}(t) \tilde{v}_{r_1}(t) \tilde{v}_{r_2}(t) \tilde{v}_{r_3}(t) \right|}.
\end{align}
The typical velocity limit becomes zero for the serious adiabaticity breakdown, while it remains finite for the non-serious adiabaticity breakdown.

\section{Velocity limit in wide-band limit}
\label{sec:VelocityLimitInWidebandLimit}

\begin{figure}[tb]
    \centering
    \includegraphics[width=0.8\columnwidth]{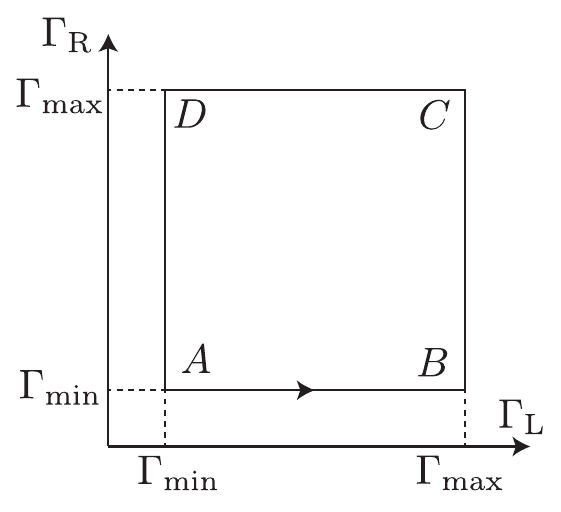}
    \caption{Square contour.}
    \label{fig:SquareContour}
\end{figure}

Here, we present the velocity limit calculated by the second-order correction and the third-order correction in the wide-band limit.
First, we consider the quantum dot system with a finite electrochemical potential bias and a square driving contour, as illustrated in Fig.~\ref{fig:SquareContour}.
Next, we estimate the electrochemical potential bias dependence for a typical velocity limit.

\subsection{Velocity limit with finite electrochemical potential bias}
\label{sec:TypicalVelocityLimit}

\begin{figure}[tb]
    \centering
    \includegraphics[width=0.9\columnwidth]{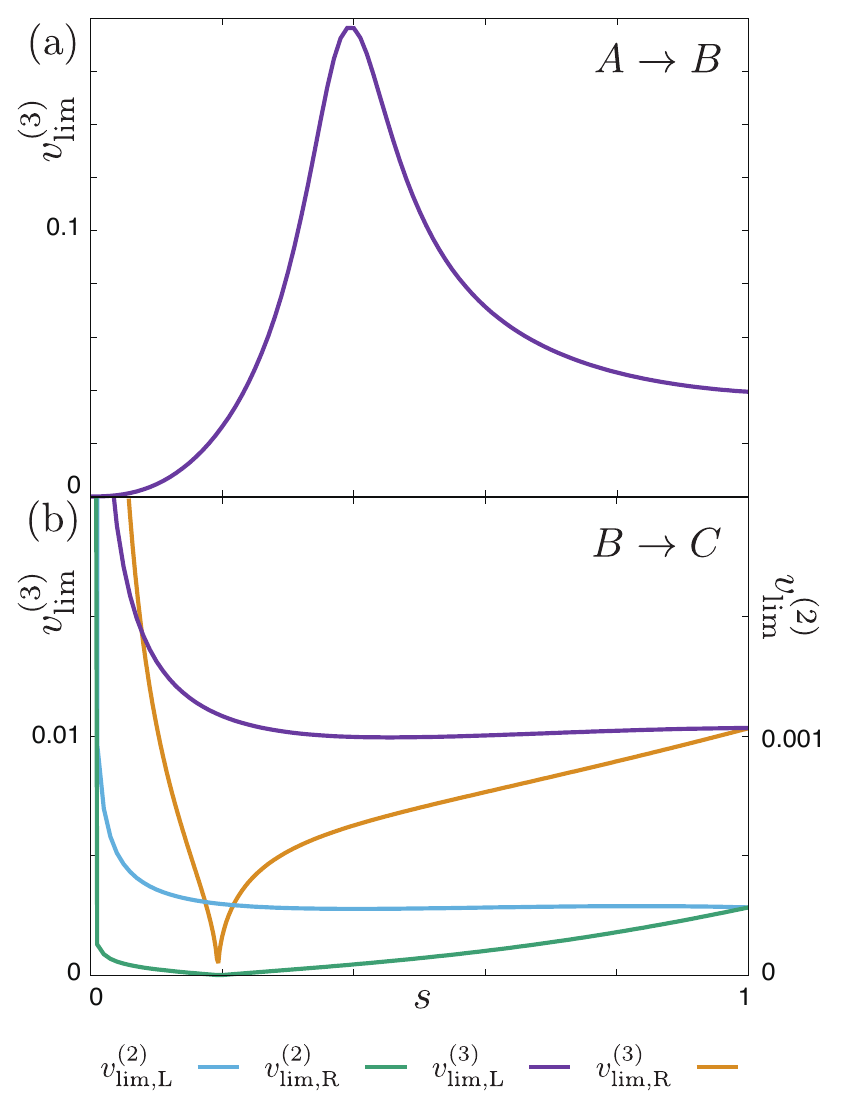}
    \caption{(Color online) Velocity limit on the line (a) $A \to B$ and (b) $B \to C$.
    A finite electrochemical potential bias is considered, $\mu_L=-\mu_R=\mu$, and $\mu$ is taken to be the unit of energy.
    The rest of the parameters are set as follows: $\epsilon_d = 0$, $\Gamma_{\mathrm{min}}=0$, $\Gamma_{\mathrm{max}}/\mu = 1.0$, $\eta=0.01$.
    The vertical axis denotes the velocity limit in units of $\mu^2$.
    $v_{\mathrm{lim,L}}^{(2)}$ and $v_{\mathrm{lim,R}}^{(2)}$ are plotted ten times larger in order to draw the lines on the same graph.
    }
    \label{fig:LocalVel}
\end{figure}

Figure~\ref{fig:LocalVel} shows four velocity limits, $v_{\mathrm{lim,L}}^{(2)}$, $v_{\mathrm{lim,R}}^{(2)}$, $v_{\mathrm{lim,L}}^{(3)}$, and $v_{\mathrm{lim,R}}^{(3)}$ on the contour shown in Fig.~\ref{fig:SquareContour}.
The limits are for the symmetric bias case, with $\mu_L = -\mu_R =\mu>0$, and $\mu$ is taken to be the unit of energy.
The parameters of the contour, $\Gamma_{\mathrm{min}}$ and $\Gamma_{\mathrm{max}}$, are taken as 0 and $\mu$, respectively (see Fig.~\ref{fig:SquareContour}).
The velocity limits are plotted for (a) line $A \to B$ and (b) line $B \to C$.
The velocity limits on lines $C \to D$ and $D \to A$ are not presented because they can be obtained by replacing the reservoir indexes ($\mathrm{L} \leftrightarrow \mathrm{R}$) in Fig.~\ref{fig:LocalVel}.
The horizontal axis denotes a dimensionless parameter $s \in [0,1]$, 
\begin{align}
    &(\Gamma_L,\Gamma_R) \nonumber \\
    &= \left\{
    \begin{array}{cc}
        s(\Gamma_{\mathrm{max}},\Gamma_{\mathrm{min}}) + (1-s) (\Gamma_{\mathrm{min}},\Gamma_{\mathrm{min}}), & (A \to B), \\
        s(\Gamma_{\mathrm{max}},\Gamma_{\mathrm{max}}) + (1-s) (\Gamma_{\mathrm{max}},\Gamma_{\mathrm{min}}), & (B \to C), 
    \end{array} \right. 
\end{align}

Only $v_{\mathrm{lim,L}}^{(3)}$ is plotted in Fig.~\ref{fig:LocalVel}~(a).
On this line, we obtain $v_{\mathrm{lim,R}}^{(2)}, v_{\mathrm{lim,R}}^{(3)} = \infty$ because reservoir R is always disconnected.
Also, when only one reservoir is connected to the dot, $v_{\mathrm{lim,L}}^{(2)}$ is infinite due to the disappearance of the second-order correction discussed in Sect.~\ref{sec:The2ndOrderCorrectionDisappearance}.
At the origin ($s=0$), the velocity limit drops to zero.
This is because the lifetime of the electron state in the quantum dot diverges and violates adiabaticity as the quantum dot becomes isolated from the reservoirs.
The peak at $s \simeq 0.4$ reflects the local maximum of the adiabatic pumping term $A_{\mathrm{L,L}}$.

In Fig.~\ref{fig:LocalVel}~(b), reservoir R is connected to the dot, so all $v_{\mathrm{lim,L}}^{(2)}$, $v_{\mathrm{lim,R}}^{(2)}$, $v_{\mathrm{lim,L}}^{(3)}$, and $v_{\mathrm{lim,R}}^{(3)}$ are plotted.
As shown, $v_{\mathrm{lim,L}}^{(2)}$ and $v_{\mathrm{lim,R}}^{(2)}$ are respectively smaller than $v_{\mathrm{lim,L}}^{(3)}$ and $v_{\mathrm{lim,R}}^{(3)}$.
At the origin ($s=0$), all the velocity limits diverge for the following reason:
around the origin, reservoir R is almost disconnected from the dot while reservoir L is strongly connected so that a small change in $\gamma_R$ does not affect charge transport at all.
This indicates that both the adiabatic pumping term and the other nonadiabatic corrections are almost zero.
The divergence occurs because the nonadiabatic corrections more rapidly decrease to zero than the adiabatic pumping term.

As shown in Fig.~\ref{fig:LocalVel}~(b), $v_{\mathrm{lim,R}}^{(2)}$ and $v_{\mathrm{lim,R}}^{(3)}$ drop to zero around $s \simeq 0.2$.
This drop happens just because the adiabatic pumping term accidentally becomes zero at this point.
This is different from the decrease in the velocity limit brought about by the isolation of the quantum dot because the nonadiabatic corrections do not diverge and remain finite.
In terms of the velocity limit, the decrease of velocity limit toward $(\Gamma_{\mathrm{L}},\Gamma_{\mathrm{R}}) = (0,0)$ shown in Fig.~\ref{fig:LocalVel}~(a) should be more serious than this drop in Fig.~\ref{fig:LocalVel}~(b), because the nonadiabatic corrections diverge in the former case.

\begin{figure}[tb]
    \centering
    \includegraphics[width=0.9\columnwidth]{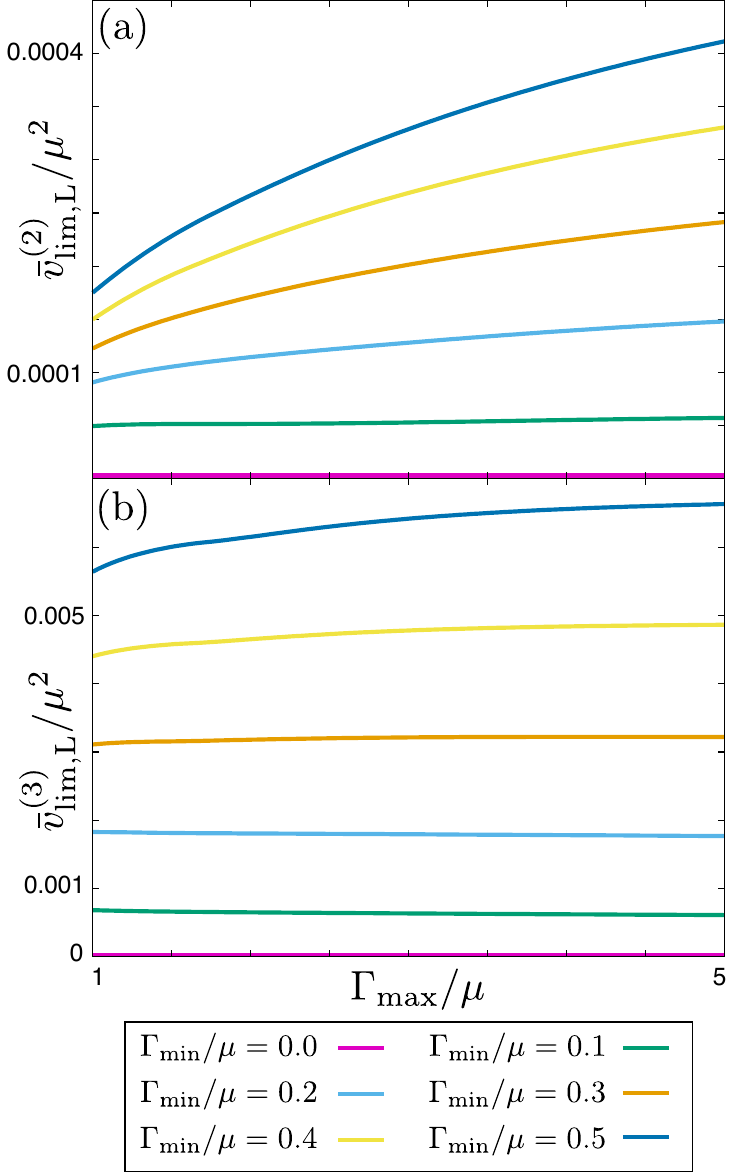}
    \caption{(Color online) Typical velocity limit calculated by (a) the second-order correction and (b) the third-order correction.
    The electrochemical potential bias is symmetric, $\mu_{\mathrm{L}}=-\mu_{\mathrm{R}}=\mu>0$. The other parameters are set as, $\epsilon_d = 0$ and $\eta = 0.01$.
    All quantities are plotted in energy units of $\mu$.
    }
    \label{fig:avevel}
\end{figure}

Now let us examine the typical velocity limit defined in Sect.~\ref{sec:TypicalVelocityLimit}.
It is plotted in Figure~\ref{fig:avevel}~(a) $\bar{v}_{\mathrm{lim,L}}^{(2)}$ and (b) $\bar{v}_{\mathrm{lim,L}}^{(3)}$ on different square contours.
We assume a symmetric electrochemical potential bias, i.e., $\mu_{\mathrm{L}}=-\mu_{\mathrm{R}}=\mu>0$.
Six cases are plotted: $\Gamma_{\mathrm{min}} / \mu = 0,0.1,0.2,0.3,0.4,0.5$.
As shown in Figs.~\ref{fig:avevel}~(a) and (b), the velocity limit decreases to zero and becomes insensitive to $\Gamma_{\mathrm{max}}$ as the driving contour reaches the origin $(\Gamma_{\mathrm{L}},\Gamma_{\mathrm{R}}) = (0,0)$.
This reflects the isolation of the quantum dot.
Around the point $(\Gamma_{\mathrm{L}},\Gamma_{\mathrm{R}}) = (0,0)$, due to the isolation effect, the nonadiabatic corrections diverge, so that the typical velocity limit strongly depends on the speed limit around the origin.
On the other hand, once the driving contour is far from the origin, nonadiabatic corrections remain finite so that the typical velocity limit becomes sensitive to $\Gamma_{\mathrm{max}}$.
Also, comparing Fig.~\ref{fig:avevel}~(a) with Fig.~\ref{fig:avevel}~(b), we can see that $\bar{v}_{\mathrm{lim,L}}^{(2)}$ is almost ten times smaller than $\bar{v}_{\mathrm{lim,L}}^{(3)}$.
This supports our conclusion that the velocity limit is imposed by the second-order correction in most cases.

\subsection{Electrochemical potential bias dependence of velocity limit}

\begin{figure}[tb]
    \centering
    \includegraphics[width=0.9\columnwidth]{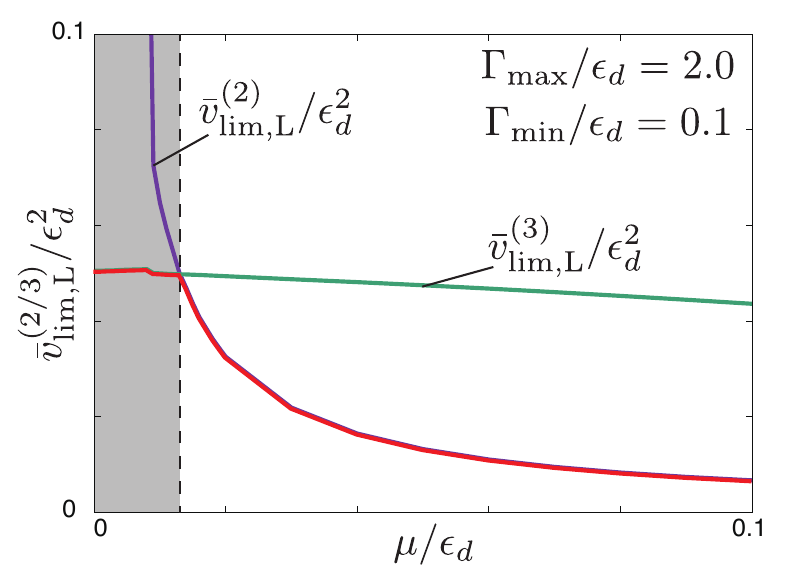}
    \caption{(Color online) Electrochemical potential bias dependence of typical velocity limits, $\bar{v}_{\mathrm{lim,L}}^{(2)}$ (purple) and $\bar{v}_{\mathrm{lim,L}}^{(3)}$ (green).
    The red line shows the minimum value between $\bar{v}_{\mathrm{lim,L}}^{(2)}$ and $\bar{v}_{\mathrm{lim,L}}^{(3)}$.
    The electrochemical potential bias is $\mu_{\mathrm{L}}=-\mu_{\mathrm{R}}=\mu>0$.
    The other parameters are set as $\Gamma_{\mathrm{max}} / \epsilon_d = 2.0$, $\Gamma_{\mathrm{max}} / \epsilon_d = 0.1$, and $\eta = 0.01$.
    All quantities are plotted in energy units of $\epsilon_d$.
    }
    \label{fig:VarMu}
\end{figure}

The previous section showed that the velocity limit is mostly determined by the second-order correction, but there is an exception as discussed in Sect.~\ref{sec:The2ndOrderCorrectionDisappearance};
i.e., when the electrochemical potential bias is zero, the second-order correction disappears and the typical velocity limit $\bar{v}_{\mathrm{lim,L}}^{(2)}$ diverges.
However, for a finite electrochemical potential bias, the second-order correction remains finite and $\bar{v}_{\mathrm{lim,L}}^{(2)}$ becomes smaller than $\bar{v}_{\mathrm{lim,L}}^{(3)}$.
Thus, it is expected that the velocity limit should be determined by $\bar{v}_{\mathrm{lim,L}}^{(3)}$ for small electrochemical potential biases, while it should be determined by $\bar{v}_{\mathrm{lim,L}}^{(2)}$ for large electrochemical potential biases.

To see this transition, let us examine the electrochemical potential bias dependence of the typical velocity limit.
Figure~\ref{fig:VarMu} shows the velocity limits, $\bar{v}_{\mathrm{lim,L}}^{(2)}$ (purple line) and $\bar{v}_{\mathrm{lim,L}}^{(3)}$ (green line).
The red line plots $\mathrm{min}[ \bar{v}_{\mathrm{lim,L}}^{(2)}, \bar{v}_{\mathrm{lim,L}}^{(3)}]$. 
Let us consider a fixed square contour, $\Gamma_{\mathrm{max}}/\epsilon_d=1.0$ and $\Gamma_{\mathrm{min}}/\epsilon_d=0.1$, and estimate typical velocity limits, $\bar{v}_{\mathrm{lim,L}}^{(2)}$ and $\bar{v}_{\mathrm{lim,L}}^{(3)}$, for different electrochemical potential biases, $\mu_{\mathrm{L}}=-\mu_{\mathrm{R}}=\mu>0$.
Different from Fig.~\ref{fig:avevel}, we will set the dot level $\epsilon_d \neq 0$ and employ it as the energy unit.
This is because we want to avoid the special point, $\epsilon_d = \mu = 0$, where no charge can be pumped so that all the higher-order corrections are zero.

As shown in the figure, for small biases, $\bar{v}_{\mathrm{lim,L}}^{(2)}$ diverges, while  $\bar{v}_{\mathrm{lim,L}}^{(3)}$ remains finite.
For large biases, $\bar{v}_{\mathrm{lim,L}}^{(2)}$ converges to a finite value which is sufficiently smaller than $\bar{v}_{\mathrm{lim,L}}^{(3)}$.
This result indicates that the velocity limit should be estimated by $\bar{v}_{\mathrm{lim,L}}^{(2)}$ when the electrochemical potential bias is finite, while it should be estimated by $\bar{v}_{\mathrm{lim,L}}^{(3)}$ when the bias is small.

\section{Adiabaticity breakdown in band-edge system}
\label{sec:AdiabaticityBreakDownInBandEdgeSystem}

When the system shows critical phenomena, for example, phase transition, transport changes drastically at the critical point and the adiabaticity easily breaks down in this regime.
This adiabaticity breakdown can be observed as a divergence of nonadiabatic corrections.
In this section, we examine this adiabaticity breakdown in a quantum dot system with the electron reservoirs that have a band edge.
There are two cases in which adiabaticity breaks down:
(i) the quantum dot is isolated from all the reservoirs, $\gamma_L = \gamma_R = 0$, so  adiabaticity trivially breaks down because the quantum dot never relaxes;
(ii) when the dot level is located near the band edge of the electron reservoirs, a bound state with an infinite lifetime appears in the quantum dot system and it violates adiabaticity.

To see the details, let us assume the reservois have the following density of states:
\begin{align}
    \rho(\omega) = \Theta (\omega) \rho_{0} \left( \frac{\omega}{\omega_c} \right)^{\alpha} \exp \left[- \frac{\omega}{\omega_c} \right]
\end{align}
Here, we assume that the band edge is at $\omega=0$ and $\omega_c$ and $\rho_{0}$ are the cutoff frequency and amplitude of the density of states, respectively.
The bound state appears when the coupling constants between the dot and the reservoirs are larger than a critical value:~\cite{Yang2015Oct,Jussiau2019Sep,Hasegawa2019b}
\begin{align}
    \pi \rho_{0} \sum_r |\gamma_r(t)|^2 = \frac{\epsilon_d}{\Gamma(\alpha)},
\end{align}
where $\Gamma(\alpha)$ is the Gamma function.
As a quantum state with an infinite lifetime emerges, the system cannot relax and any finite-velocity driving violates adiabaticity.
Such a violation would appear as a divergence of nonadiabatic corrections.

\begin{figure}[tb]
    \centering
    \includegraphics[width=0.9\columnwidth]{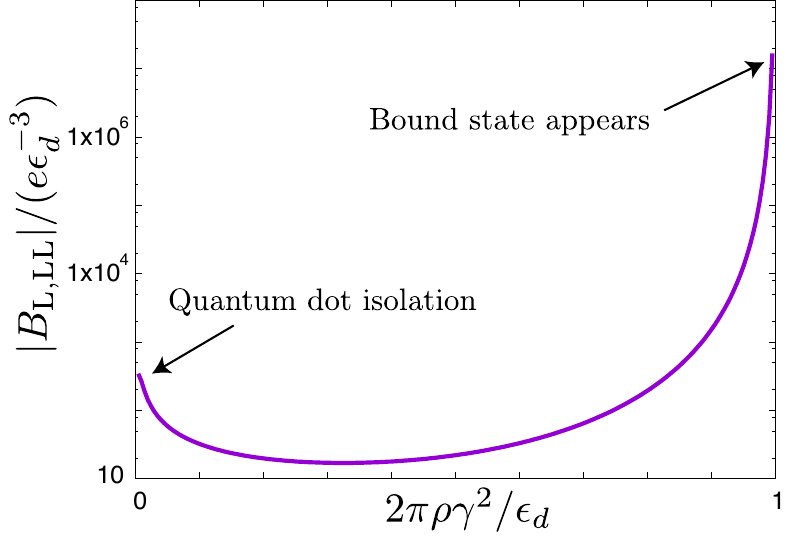}
    \caption{(Color online) Coupling strength dependence of the second-order correction, $B_{r,\mathrm{LL}}(\gamma_{\mathrm{L}},\gamma_{\mathrm{R}})$.
    The parameters are set as follows: $\mu_{\mathrm{L}}/\epsilon_d=0$, $\mu_{\mathrm{R}}/\epsilon_d = 2.0$, $\omega_c /\epsilon_d = 10.0$, $\alpha = 1.0$.
    The vertical axis is a logarithmic scale.
    }
    \label{fig:BE_B}
\end{figure}
To see this, let us consider small $\gamma_L$ driving,
\begin{align}
(\gamma_{\mathrm{L}},\gamma_{\mathrm{R}}) \to (\gamma_{\mathrm{L}}+\Delta \gamma,\gamma_{\mathrm{R}})
\end{align}
over an elapsed time $\Delta t$.
The second-order correction for this driving is calculated as
\begin{align}
    \delta Q_r^{(2)} = B_{r,\mathrm{LL}}(\gamma_{\mathrm{L}},\gamma_{\mathrm{R}})  \frac{(\Delta \gamma)^2}{\Delta t}  .
\end{align}
Figure \ref{fig:BE_B} presents coupling constant dependence of $B_{r,\mathrm{LL}}(\gamma_{\mathrm{L}},\gamma_{\mathrm{R}})$.
A symmetric coupling, $\gamma_{\mathrm{L}}=\gamma_{\mathrm{R}}=\gamma$, is considered.
As shown in the figure, there are two divergences: the one at $2\pi \rho_0 \gamma^2 / \epsilon_d= 0$ corresponds to the isolation of the quantum dot and the one at $2\pi \rho_0 \gamma^2 / \epsilon_d= 1.0$ corresponds to the emergence of the bound state.
The singularity at $2\pi \rho_0 \gamma^2 / \epsilon_d= 1.0$ is much stronger than that at $2\pi \rho_0 \gamma^2 / \epsilon_d= 0$.
This indicates that the emergence of the bound state with an infinite lifetime is fatal to adiabaticity and adiabatic operations cannot be performed around this critical point.

\section{Summary}
\label{sec:Summary}

We studied nonadiabatic corrections in charge pumping and derived a formula for adiabatic criteria by employing a noninteracting electron model of a single-level quantum dot as a minimum model.
The upper velocity limit for the parameter for adiabatic pumping was derived from the $n$th-order term of the pumped charge, which is proportional to $T^{n-1}$ where $T$ is the period of the pumping cycle.
The leading nonadiabatic correction derived from the second term ($n=2$) imposes the strongest constraint in most cases.
However, the next leading correction derived from the third term ($n=3$) becomes important when the two reservoirs have the same electrochemical potential and the same temperature because the leading nonadiabatic correction vanishes in such a situation.
As a demonstration, we considered charge pumping induced by the time-dependent dot-reservoir coupling and calculated the velocity limit in the wide-band limit.
We found that the velocity limit indeed depends on the point in the parameter space; this feature is important for optimization to maximize the pumping power.
We also showed how the velocity limits imposed from the second- and third-order terms behave near thermal equilibrium.

To detect adiabatic breakdown, we defined a typical velocity limit by averaging the velocity limit on the parameter contour.
We showed that the typical velocity limit approaches zero 
in the wide-band limit when the dot-reservoir coupling is weakened.
We also calculated the typical velocity limit for a quantum dot coupled to reservoirs with a band edge and showed that it approaches zero, corresponding to adiabatic breakdown caused by the appearance of a bound state in this model.

We expect that our findings will be helpful not only for accurate control of adiabatic pumping in quantum devices but also for the foundation of non-equilibrium statistics such as in the study of quantum heat engines.
While our work illustrates a general procedure for obtaining adiabatic criteria just in a simple model, it is a straightforward exercise to extend our formulation to more complex models. 
An important future problem would be to extend our work to take the effect of Coulomb interactions between electrons into account.

\begin{acknowledgment}
M.H. acknowledges the financial support provided by the Advanced Leading Graduate Course for Photon Science.
T.K. was supported by JSPS Grants-in-Aid for Scientific Research (No. JP24540316, JP26220711, and JP20K03831).
\end{acknowledgment}

\appendix

\section{Keldysh Green's functions}
\label{sec:KeldyshGreensFunction}

In this appendix, we define the Keldysh Green's functions (GFs) and the one-particle-irreducible self-energies.
The retarded and advanced GFs of an electron in the isolated quantum dot are defined as
\begin{align}
    g^{R}(t_1,t_2) &= -i \Theta(t_1-t_2) \Braket{[d(t_1),d^{\dagger}(t_2)]_{+}}_{\gamma_r=0}, \\
    g^{A}(t_1,t_2) &= i \Theta(t_2-t_1) \Braket{[d(t_1),d^{\dagger}(t_2)]_{+}}_{\gamma_r=0},
\end{align}
and those of electrons in the reservoir $r$ with wavenumber $k$ are defined as
\begin{align}
    g_{rk}^{R}(t_1,t_2) &= -i \Theta(t_1-t_2) \Braket{[c_{rk}(t_1),c_{rk}^{\dagger}(t_2)]_{+}}_{\gamma_r=0}, \\
    g_{rk}^{A}(t_1,t_2) &= i \Theta(t_2-t_1) \Braket{[c_{rk}(t_1),c_{rk}^{\dagger}(t_2)]_{+}}_{\gamma_r=0}.
\end{align}
Here $[\cdot,\cdot]_+$ is the anti-commutator, and $\braket{\cdot}$ denotes the ensemble average.
The retarded GF of a dot electron for a finite dot-reservoir coupling is calculated using the Dyson equation:
\begin{align}
    &G^{R}(t_1,t_2) = g^{R}(t_1,t_2) \nonumber \\
    &\hspace{0.5cm} + \! \int \! dt_3 dt_4 \, g^{R}(t_1,t_3) \Sigma^{R}(t_3,t_4) G^{R}(t_4,t_2),
\end{align}
where $\Sigma^{R}(t_1,t_2)$ is the one-particle-irreducible self-energy for the retarded GF, defined as
\begin{align}
    \Sigma^{R}(t_1,t_2) &= \sum_r \Sigma_r^{R}(t_1,t_2) ,\\
    \Sigma_r^{R}(t_1,t_2) &= \sum_k \gamma_r(t_1) g_{rk}^{R}(t_1,t_2) \gamma_r(t_2) .
\end{align}
The advanced GF and its self-energy are obtained from the retarded ones as $G^A(t_1,t_2) = -G^R(t_2,t_1)$ and $\Sigma^A(t_1,t_2) = -\Sigma^R(t_2,t_1)$, respectively.
The lesser GF is calculated from the retarded and advanced GFs: 
\begin{align}
    G^<(t_1,t_2) = \! \int \! dt_3 dt_4 \, G^R(t_1,t_3) \Sigma^<(t_3,t_4) G^A(t_4,t_2) .
\end{align}
where $\Sigma^<(t_1,t_2)$ is the lesser component of the self-energy defined as
\begin{align}
    \Sigma^{<}(t_1,t_2) &= \sum_r \Sigma_r^{<}(t_1,t_2) ,\\
    \Sigma_r^{<}(t_1,t_2) &= \sum_k \gamma_r(t_1) g_{rk}^{<}(t_1,t_2) \gamma_r(t_2) .
\end{align}

\section{Adiabatic condition for the acceleration}
\label{sec:ConditionAcceleration}

We define the acceleration of the driving parameters and its normalized value as
\begin{align}
    & {\bm a}(t) = (a_{\rm L}(t),a_{\rm R}(t))
    = \biggl( \frac{d^2 \gamma_{\rm L}(t)}{d t^2}, 
    \frac{d^2 \gamma_{\rm R}(t)}{d t^2}\biggr), \\
    & \tilde{\bm a}(t) = (\tilde{a}_{\rm L}(t),\tilde{a}_{\rm L}(t))
    = \frac{{\bm a}(t)}{|{\bm a}|},
\end{align}
respectively.
The adiabatic condition, Eq.~(\ref{eq:condI3a}), can be rewritten as
\begin{align}
    & |a_{r}(t)| \le a_{\mathrm{lim},r}^{(3)}(t), \\
    & a_{\mathrm{lim},r}^{(3)} = \eta\frac{|A_{r}(t)|}{|D_r(t)|} , \\
    & D_r(t) = \sum_{r_1,r_2} D_{r,r_1r_2}(t) \tilde{a}_{r_1}(t) \tilde{v}_{r_2}(t).
\end{align}
Here $a_{\mathrm{lim},r}^{(3)}(t)$ is the limit of the acceleration of the driving parameters.

In our work, we consider the driving contour shown in Fig.~\ref{fig:SquareContour} with the protocol of constant velocity for simplicity, which leads to infinite acceleration at the corners of the driving contour.
This apparent singularity can be avoided by  implicitly assuming a finite margin near the corners where the acceleration is finite.
For this finite acceleration, the corresponding nonadiabatic correction is considered to be much smaller than other corrections by the following reasons.
For rough estimate, we can assume that $D_{r,r_1,r_2}(t)$ is the same order with $C_{r,r_1,r_2,r_3}(t)$ unless it has a singularity.
Then, by the dimensional analysis, the nonadiabatic correction by finite acceleration becomes much smaller than that by the third corrections with respect to the velocity if the time scale of the parameter driving is longer than $\Gamma_{\rm min}^{-1}$ (compare Eq.~(\ref{eqn:3rdOrderGeometric}) with Eq.~(\ref{eqn:3rdOrderGeometricv})).

The above discussion holds only when $D_{r,r_1,r_2}(t)$ has no singularity. 
If it has a singularity such as the adiabaticity breakdown, the nonadiabatic correction by finite acceleration may be relevant.
However, in such a situation, the higher order corrections also become important.
The analysis for such a situation is beyond the scope of our work and is left as a future problem.

\bibliographystyle{jpsj}
\bibliography{references}

\end{document}